\documentclass[twocolumn]{revtex4-2}

\usepackage{graphicx}
\usepackage{dcolumn}
\usepackage{bm}
\usepackage{hyperref}
\usepackage[mathlines]{lineno}
\usepackage[dvipsnames]{xcolor}
\usepackage{color}
\usepackage{amsfonts}
\usepackage{amsopn}
\usepackage{amsmath}

\newcommand{\beq}{\begin{equation}}
\newcommand{\eeq}{\end{equation}}
\newcommand{\barr}{\begin{eqnarray}}
\newcommand{\earr}{\end{eqnarray}}
\newcommand{\bseq}{\begin{subequations}}
\newcommand{\eseq}{\end{subequations}}

\newcommand{\expectation}[3]{\langle #1|#2|#3\rangle}

\newcommand{\ket}[1]{|#1\rangle}

\newcommand{\vett}[1]{\textbf{#1}}
\newcommand{\uvett}[1]{\hat{\textbf{#1}}}
\begin{document}

\preprint{APS/123-QED}

\title{The Geometry of Paraxial Vector Beams}

\author{Marco Ornigotti}
\affiliation{$^1$Tampere University, Photonics Laboratory, Physics Unit, Tampere, FI-33720, Finland}
 \email{marco.ornigotti@tuni.fi}

\date{\today}

\begin{abstract}
This work unveils a novel and fundamental connection between structured light and topological field theory by showing how the natural geometrical setting for paraxial vector beams is that of a $SU(2)$ principal bundle over $\mathbb{R}^{2+1}$. Going beyond the usual high-order Poincar\'e sphere approach, we show how the nonseparable structure of polarisation and spatial modes in vector beams is naturally described by a non-Abelian Chern-Simons gauge theory. In this framework, we link the Chern-Simons charge to spin-orbit coupling, and we propose a simple way to experimentally detect the presence of  non-Abelian phases through Wilson lines. This new insight on vector beams opens new possibilities for realising and probing topological quantum field theories using classical optics, as well as it lays the foundation for implementing topologically protected classical and quantum information protocols with structured light.
\end{abstract}

\maketitle
\emph{Introduction-} Geometry and topology form the foundational language of modern theoretical physics, offering deep insights into fundamental and emergent phenomena. From Dirac's quantisation of charge \cite{diracMonopole} to the classification of topological phases of matter \cite{topInsulatorBook}, geometry and topology underpin a wide variety of physical theories. The development of topological quantum field theories (TQFTs), for example, has led to new discoveries in high-energy and condensed matter physics, including the discovery of anyons \cite{anyonsBook}, the theoretical and experimental understanding of the quantum \cite{QHeffect} and fractional \cite{FQHeffect} Hall effects, and the emergence of topologically protected states with non-Abelian statistics \cite{NAcomp}. Witten’s seminal work linking TQFT and knot theory \cite{witten1}, together with the formalism of modular tensor categories \cite{MTCref}, laid the theoretical foundation for topological quantum computing (TQC) \cite{TQCbook}, where information is encoded in the global, nonlocal properties of anyonic systems \cite{anyons1, anyonInterferometry}. These ideas have 
since influenced other fields of physics, such as fluid dynamics, where helicity and vortex lines have been reinterpreted through the lens of knot invariants and field topology \cite{fluids2}, photonics, with the advent of topological photonics \cite{topPhotonics}, atomic physics through synthetic gauge field engineering in optical lattices \cite{TopAtom}, and electronics, with the discovery of topoelectrical circuits \cite{TopElectric}.

Within this context structured light (SL) offers a particularly promising platform for testing topological and geometrical concepts, while at the same time drawing new ways to control and shape the flow of light from them. From the seminal works of Berry and Nye on wave dislocations in 1974 \cite{berryNye}, and Allen and Woerdman on Laguerre-Gaussian beams carrying orbital angular momentum (OAM) in 1992 \cite{allenWoerdman}, the field of SL has rapidly advanced, enabling a wide range of applications in microscopy \cite{ref1}, spectroscopy \cite{ref2}, sensing \cite{ref20}, and both classical and quantum information processing \cite{ref4,ref5,ref7}. Alongside these applications, tools from topology \cite{ref8, ref9}, knot theory \cite{isolatedKnots,ref13,ref13bis}, differential geometry \cite{ref10,ref11}, have been employed to further understand the structure of light field, establishing SL as a natural interface between field theory and experimental optics.

Of particularly interest within SL are vector beams (VBs), i.e., paraxial optical fields with spatially varying polarisation\cite{ragazzina1,ragazzina2,milione}, since they possess particularly rich geometrical and topological features due to their intrinsic spin-orbit coupling (SOC), manifesting as nonseparable correlations between spatial and polarisation degrees of freedom \cite{ref14,ref16,ref17}. This enables unique applications in metrology \cite{ref19}, polarisation-sensitive sensing \cite{ref20,lea1}, advanced material processing \cite{ref21}, and high-dimensional quantum information encoding \cite{ref22}. Beyond canonical example such as radially and azimuthally polarized beams \cite{ragazzina1,ragazzina2,milione}, more complex VBs, including Poincaré beams \cite{ref23}, skyrmionic beams \cite{ref24,ref25,ref25bis,refSkyrmions1,refSkyrmions2,refSkyrmions3, refSkyrmions4}, hopfions \cite{hopfionsMark}, optical merons \cite{meron1,meron2,meron3}, Stokes origami \cite{stokesOrigami} and, more recently, spectral VBs \cite{lea1} and  spatio-spectral VBs \cite{RobertIo}, provide good examples of the rich topological structure of VBs, and motivate a deeper investigation of their underlying geometry.

Motivated by all this, in this Letter I present a new theoretical framework, based on a non-Abelian, Chern-Simons (CS) gauge theory description of VBs,  that provides a unified perspective on the topological and geometrical properties of VBs, enabling a classification of VBs based on their topology. Specifically, I will show that VBs naturally define a $SU(2)$ principal bundle over $\mathbb{R}^{2+1}$, extending their geometrical understanding beyond the standard high-order Poincar\'e sphere (HOPS) picture \cite{milione, ragazzina1}, and providing a natural setting for the observation of non-Abelian geometric phases. In this setting, I will construct a CS connection encoding the beam's structure and use its correspondent topological invariants, such as CS charge and Wilson holonomy, to classify VBs according to their topological properties. This framework not only advances further our understanding of SL from a topological field theory perspective, but also establishes a direct link between these two disciplines, which can be used to define novel protocols for classical and quantum information processing and computation. 

\emph{Fibre Bundle Approach to VBs -} VBs are commonly defined as bipartite states of polarisation and orbital angular momentum (OAM), spanning a four-dimensional Hilbert space $\mathcal{H}_4=\text{span}\{\ket{L,\ell},\ket{R,m},\ket{L,-m},\ket{R,-\ell}\}$, where $\{\ket{L},\ket{R}\}$ indicate left- and right-handed circular polaristion states, respectively, and $\{\ket{\ell},\ket{m}\}$ are Laguerre-Gaussian modes of the form $LG_0^{\ell}(r,\varphi,z)=f_{|\ell|}(r,z)\exp(i\ell\varphi)\exp[i\zeta_{\ell}(z)]$, with $f_{|\ell|}(r,z)$ accounting for their radial structure, $\ell$ being their OAM, and $\zeta_{\ell}(z)$ the Gouy phase. This space admits a natural factorisation $\mathcal{H}_4=\{\ket{L,\ell},\ket{R,m}\}\oplus\{\ket{L,-m},\ket{R,-\ell}\}\equiv\mathcal{H}_+\oplus\mathcal{H}_-$, where each subspace can be mapped into a HOPS \cite{milione, ragazzina1,ragazzina2, hybridPoincare}.

For example, a general beam in $\mathcal{H}_+$ takes the form $\ket{E_+}=\cos(\theta/2)\ket{L,\ell}+\sin(\theta/2)\exp(i\phi)\ket{R,m}$, where $\{\theta,\phi\}$ parametrise the HOPS. Such a beam carries a Berry connection $A^{+}_{\mu}=-i\,E_{+}^{\dagger}\,dE_{+}$, whose components can be written in terms of the Stokes vector $\vett{n}_+=\expectation{E_{+}}{\boldsymbol\sigma}{E_{+}}$, recovering the familiar $U(1)$ bundle picture of polarisation geometry.

While simple, this framework only captures beams confined to $\mathcal{H}_{\pm}$, whose geometry reduces to that of their respective HOPS. Beams that weave across both subspaces, or span a different combination of polarisation and OAM, admit a richer geometric structure that is not generally captures by the HOPS picture. In such cases, in fact, the local mode-polarisation frame can rotate nontrivially in $SU(2)$, resulting in matrix-valued geometric phases beyond Berry's Abelian form. These situations are more naturally described as $SU(2)$ principal bundles over $\mathbb{R}^{2+1}$, where the ``$+1$" denotes the propagation coordinate. In this non-Abelian setting, the fibre bundle captures the full spin-orbit structure of the beam, allowing for topological features that cannot be faithfully be represented on a single Poincar\'e sphere (PS).
\begin{figure*}
    \begin{center}
        \includegraphics[width=\textwidth]{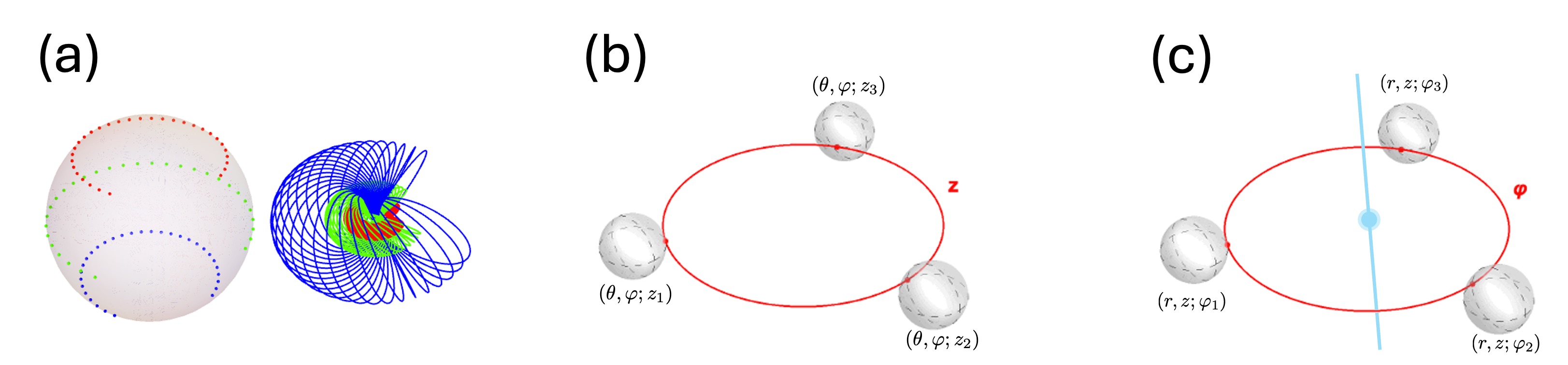}
        \caption{Pictorial representation of the possible choices of compactification for a paraxial VB. (a) One point compactification at infinity ($\mathbb{R}^{2+1}\cup\{\infty\}\cong S^3$. This is the most common compactification strategy, resulting in identifying the beam with $S^3$, here represented through its Hopf fibration over the Poincar\'e sphere, where each point on the sphere corresponds to a great circle on $S^3$ \cite{nakahara, mandelWolf}. (b) Rayleigh range compactification ($\mathbb{R}^{2+1}\cong S^2\times S^1$. Here, the $z$-direction is compactified into a circle by identifying the endpoints of the propagation interval $[-nz_R,nz_R]$. At each point over the circle $S^1$ (red line, labelled by $z$) we define a Poincar\'e sphere, identified using the spherical angles $(x,y)\cong(\theta,\varphi)$ using one point compactification at infinity on the transverse plane. (c) Vortex compactification ($\mathbb{R}^{2+1}\cong S^2\times S^1$). This procedure is similar to that in (b), but here we compactify the azimuthal direction $\varphi$ into the circle $S^1$ (red line, labelled $\varphi$) by identifying the endpoints of the interval $[0,2\pi]$. This results in a circle $S^1$ threading a vortex line (blue line), corresponding to the phase singularity carried by the beam. As before, at each point along the $\varphi$-circle we associate a Poincar\'e sphere identified by the radial and propagation coordinates $(r,z)$ of the beam, using one point compactification on the $(r,z)-$plane.}
        \label{figure2}
    \end{center}
\end{figure*}

We refer to this $SU(2)$ principal bundle $\mathcal{P}(\mathbb{R}^{2+1},SU(2),\pi)$ as the \emph{paraxial bundle}. Its base space, $\mathbb{R}^{2+1}$, represents the transverse plane of the beam, together with its propagation direction. The fibre is the group $SU(2)$, acting on the polarisation degree of freedom, and the projection $\pi:\mathcal{P}\rightarrow\mathbb{R}^{2+1}$ assigns each group element to its spatial location, i.e., to a particular OAM component of the beam. 
%

A normalised spinor field $\eta=a\ket{\Psi(\vett{r})}+b\ket{\Phi(\vett{r})}$, with $\ket{\Psi(\vett{r})}$, $\ket{\Phi(\vett{r})}$ two orthogonal vectors in $\mathcal{H}$, defines a local section of the bundle, i.e., a smoothly varying choice of mode-polarisation configuration at each point in space. Local gauge transformations $\eta'=g\eta$, with $g:\,\mathbb{R}^{2+1}\,\rightarrow\,SU(2)$, act as spatially dependent polarisation rotations, internal or external, linking the description of the beam at different points. These transformations naturally generate a non-Abelian connection $M=\eta^{\dagger}d\eta$, which transforms under $SU(2)$ as $M'=g^{\dagger}\,M\,g-g^{\dagger}dg$ \cite{nakahara, gockeller}. The curvature $F=dM+M\wedge M$ of this connection encodes the beam's spin-orbit topology, and gauge-invariant quantities derived from it, such as CS charges and non-Abelian holonomies, classify distinct topological sectors of VB configurations.
%

\emph{Full Non-Abelian Structure of VBs -}  To reveal the complete $SU(2)$ gauge geometry of the paraxial bundle, we begin by choosing a convenient local frame $\{\ket{\Psi(\vett{r})},\ket{\Phi(\vett{r})}\}\in\mathcal{H}_4$. Without loss of generality, let us $\ket{\Psi(\vett{r})}$ lie on the PS spanned by $\{\ket{L,\ell},\ket{R,m}\}$, and $\ket{\Phi(\vett{r})}$ on the one spanned by $\{\ket{L,-m},\ket{R,-\ell}\}$, i.e.,
%
\bseq\label{eqs2}
\begin{align}
\ket{\Psi(\vett{r})}&=\left(\begin{array}{c}
\cos\frac{\theta(r,z)}{2}\,e^{i\phi_{\ell}(\varphi,z)}\\
\sin\frac{\theta(r,z)}{2}\,e^{i\phi_m(\varphi,z)}
\end{array}\right),\\
\ket{\Phi(\vett{r})}&=\left(\begin{array}{c}
-\sin\frac{\theta(r,z)}{2}\,e^{-i\phi_m(\varphi,z)}\\
\cos\frac{\theta(r,z)}{2}\,e^{-i\phi_{\ell}(\varphi,z)}
\end{array}\right),
\end{align}
\eseq
where $\tan\theta(r,z)=f_{|m|}(r,z)/f_{|\ell|}(r,z)$, and $\phi_{\ell}(\varphi,z)=\ell\varphi+\psi_{\ell}(z)$. These spinors, associated with distinct local PSs, define a polarisation-OAM frame. From these, one can assemble a section $U:\,\mathbb{R}^{2+1}\,\rightarrow\,\mathcal{P}$ of the paraxial bundle in the form of an $SU(2)$ matrix as follows \cite{nakahara}
\beq\label{eq3}
U(\vett{r})=\left(\begin{array}{cc}
\cos\theta(r,z)\,e^{i\phi_{\ell}(\varphi,z)} & -\sin\theta(r,z)\,e^{-i\phi_m(\varphi,z)}\\
\sin\theta(r,z)\,e^{i\phi_m(\varphi,z)} & \cos\theta(r,z)\,e^{-i\phi_{\ell}(\varphi,z)} 
\end{array}\right).
\eeq
This defines a local $SU(2)$ frame for the VB. The corresponding $\mathfrak{su}(2)$-valued connection $M=U^{\dagger}(\vett{r})dU(\vett{r})=M_{\mu}dx^{\mu}$ (with $x^{\mu}=\{r,\varphi,z\}$) encodes the full spin-orbit geometry of the beam. Unlike the Berry connection $A^{\pm}$ of the HOPS, which tracks a single $U(1)$ phase along a fixed direction in state space, $M$ retains the entire $SU(2)$ fibre structure. This richer connection captures therefore polarisation topologies beyond Abelian projection. Using the relation $M=M_{\mu}^aT^adx^{\mu}$ \cite{gockeller},  where $T^a=-i\sigma^a/2$ are the (anti-Hermitian) generators of $SU(2)$ \cite{nakahara}, we can represent it instead in terms of its components as \cite{supplementary}
\bseq\label{eqs5}
\begin{align}
M_{\mu}^1&=\partial_{\mu}\theta\sin\Delta\phi^++\sin\theta\partial_{\mu}\Delta\phi^-\cos\Delta\phi^+,\\
M_{\mu}^2&=2\partial_{\mu}\theta\cos\Delta\phi^++\sin\theta\partial_{\mu}\Delta\phi^-\sin\Delta\phi^+,\\
M_{\mu}^3&=-\left(\partial_{\mu}\Delta\phi^++\cos\theta\partial_{\mu}\Delta\phi^-\right),
\end{align}
\eseq
where $\Delta\phi^{\pm}=\phi_{\ell}\pm\phi_m$. The connection $M_{\mu}$ records both the familiar Berry phases (in its diagonal components) and the off-diagonal couplings arising from intrinsic SOC. 

This is the first result of this work: by avoiding the conventional definition of HOPS, we uncover an inherent $SU(2)$ gauge structure for VBs, arising naturally from their spatially varying polarisaiton, and providing the geometric framework to characterise their full topology and transformations.


\emph{CS Action and Topological Classification of VBs} The full topological character of a VB can be determined from the CS action
\beq\label{eqCS}
\mathcal{S}_V[M]=\frac{k}{4\pi}\int_{\mathcal{M}}\,\operatorname{Tr}\left(M\wedge dM+\frac{2}{3}M\wedge M\wedge M\right),
\eeq
where $k\in\mathbb{Z}$ is the CS level \cite{CSnotes}), the trace is taken in the fundamental representation of $SU(2)$, $\wedge$ indicates the wedge products between differential forms \cite{nakahara}, and $\mathcal{M}$ denotes a compactification of the beam's base space $\mathbb{R}^{2+1}$. As in all CS theory, such a compactification is essential if the action is to define a genuine topological invariant \cite{CSnotes, CSnotes2}.

The most familiar construction is the hypersphere compactification, in which all points at infinity are identified so that $\mathbb{R}^{2+1}\cup\{\infty\}\cong S^3$, underlying the conventional PS picture \cite{mandelWolf}. For paraxial beams, however, two alternative procedures arise more naturally. One, that we name \emph{Rayleigh range compactification}, exploits the finite Rayleigh range, i.e., the fact that beyond a few Rayleigh lengths the field intensity becomes negligible, so the physically relevant propagation interval can be closed into a loop, compactifying $z$ to a circle $S^1$, while the transverse $(r,\varphi)$ plane is one-point compactified to $S^2$, i.e., the PS. The base space is then $S^2\times S^1$, a topology particularly well-suited to describing beams with knotted structure \cite{isolatedKnots}. The other compactification procedure, which we name \emph{vortex compactification}, is generated by OAM: here the azimuthal coordinate $\varphi$ is compactified to $S^1$, while the remaning $(r,z)$ coordinates form the compactified sphere $S^2$. Again, the result is $S^2\times S^1$, but now the circle is threaded by the OAM vortex singularity. Although these two constructions lead to the same product topology, they differ in which coordinates form the circle and which form the sphere, and so yield distinct interpretations of the winding and its physical origin. Unlike the case of $S^3$, both of these compactifications encode an intrinsic, nontrivial topology tied directly to the structure of the beam. A pictorial representation of these various compactification procedures is given in Fig. \ref{figure2}.  

For a VB described by Eqs. \eqref{eqs2} the resulting CS action evaluates to, using vortex compactification \cite{supplementary},
\beq\label{eq6}
\mathcal{S}_{V}[M]=2\pi\,k\,\operatorname{deg}(U),
\eeq
where $\operatorname{deg}(U)$ is the degree of the map $U:\mathbb{R}^{2+1}\rightarrow\mathcal{P}$ defined in Eq. \eqref{eq3}, i.e., the number of times $U$ winds the sphere $S^2$ \cite{nakahara}. Explicitly \cite{supplementary}
\beq\label{eq6}
\operatorname{deg}(U)=\frac{\left[m(|\ell|+1)-\ell(|m|+1)\right]}{2}.
\eeq
Since $\operatorname{deg}(U)\in\mathbb{Z}$, the result above imposes some limitations to the choice of $(\ell,m)$, and, ultimately, the possible combinations of $(\ell,m)$ leading to VBs carrying nontrivial topologies. 

This leads naturally to a classification of VBs according to whether their CS charge is trivial ($\mathcal{S}_V[M]=0$) or nontrivial ($\mathcal{S}_V[M]\neq 0$). Uniformly polarised beams, with $\ell=m$, fall into the trivial class: here, $\operatorname{deg}(U)=0$, so $\mathcal{S}_V[M])=0$. Radially and azimuthally polarised beams, corresponding to $\ell=-m$, are also topologically trivial at the CS charge level, since $\mathcal{S}_V[M]=0$ as well. In this case, in fact, the field factorises so that the connection depends only on $\varphi$, and both the term $M\wedge dM$ and the triple wedge product in Eq. \eqref{eqCS} vanish. For general $\ell\neq m$, $\operatorname{deg}(U)$ depends on whether $[m(|\ell|+1)-\ell(|m|+1)]$ is odd or even. When odd, the degree $\operatorname{deg}(U)$ necessarily evaluates to zero, again yielding a trivial topology characterised by $\mathcal{S}_V[M]=0$. When even, on the other hand, $\operatorname{deg}(U)=n$, giving $\mathcal{S}_V[M]=2\pi\,k\,n$, and the VB carries a genuine 3D nontrivial topology.
\begin{figure}
    \begin{center}
        \includegraphics[width=0.5\textwidth]{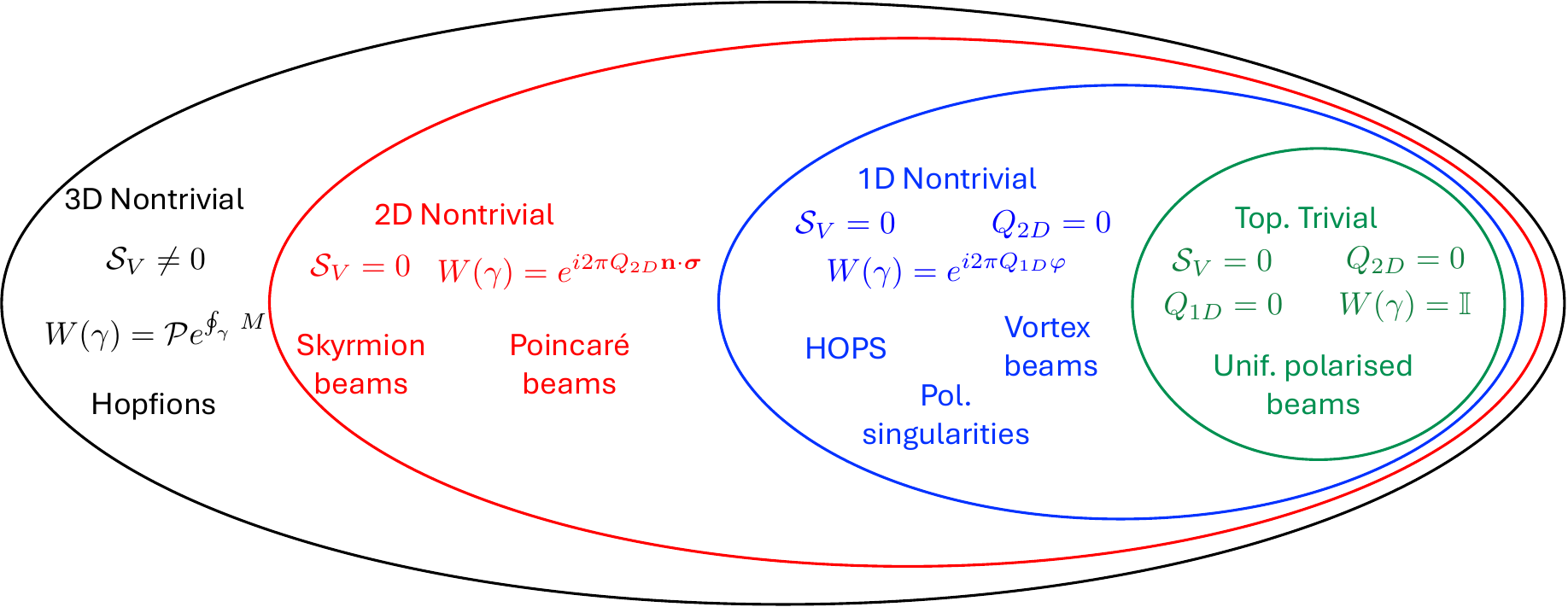}
        \caption{Pictorial representation of the topological classification of VBs, defined as in Eq. \eqref{eqs2} (a). The black ellipse contains 3D topologically nontrivial VBs, for which $\,athcal{S}_V[M]\neq 0$, and the Wilson holonomy is in general matrix-valued. An example of VBs belonging to this class are the hopfions \cite{hopfionsMark}. The red ellipse contains instead 2D topologically nontrivial VBs, possessing $\mathcal{S}_V[M]=0$, but a nonzero skyrmion-like number $Q_{2D}$. Examples of these VBs are skyrmionic beams \cite{ref24} or Poincar\'e beams \cite{ref23}. The blue ellipse contains instead the 1D topologically nontrivial VBs, for which both $\mathcal{S}_V[M]$ and $Q_{2D}$ are zero, but they possess nonzero winding number $Q_{1D}$. Representatives of this class are vortex beams, beams carrying polarisation singularities, and VBs that can be defined on a HOPS. Finally, the green ellipse contains topologically trivial VBs, characterised by all the topological charges $\mathcal{S}_V[M]$, $Q_{2D}$, and $Q_{1D}$ being zero, and hence by a trivial Wilson holonomy. This is the class, where uniformly polarised beams live.}
        \label{figure3}
    \end{center}
\end{figure}
This is the second main result of this work: the CS charge captures the full 3D topology of VBs, providing a classification invisible from the flattened perspective of the HOPS. Beams with $\mathcal{S}_V[M]\neq 0$, in fact, are not merely decorated with a surface texture, but carry a topology woven through their three-dimensional structure. Revealing this information thus requires the full non-Abelian fibre-bundle descrption, which lifts the geometry from the Abelian HOPS case to the genuinely 3D fabric encoded in the CS charge.

\emph{Wilson Holonomy and Further Classification - } The classification provided by the CS charge, however, only concerns the 3D topology of VBs. However, $\mathcal{S}_V[M]=0$ does not necessarily imply complete triviality. Paraxial skyrmionic beams ($\ell=0,\,m=1$), for example, do fall in this category, but carry nontrivial 2D topology. Analogously, a vortex beam with uniform polarisation also falls in this category, but carries nontrivial 1D topology. To complete this classification, we then need a second invariant, capable of distinguishing genuinely nontrivial VBs from those whose transverse polarisation texture carries topological structure. A good candidate for this role is the Wilson holonomy, i.e., the non-Abelian geometric phase accumulated by the connection $M$ around a closed loop $\gamma$, defined as follows
\begin{equation}\label{eqW}
W(\gamma) = \mathcal{P} \exp\left(\,\oint_{\gamma} M\,\right)=\mathcal{P}\exp\left(\int_{\Sigma}\,F\right),
\end{equation}
where $\mathcal{P}$ is the path-ordering operator, essential in this context because $M$ is generally noncommuting along the loop $\gamma$ due to its matrix-valued nature \cite{gockeller, nakahara}, $\Sigma$ is the surface bounded by $\gamma$, and $F=dM+M\wedge M$ is the curvature associated with $M$. Notice, that the line integral $\oint_{\gamma}M$ can be transformed in the surface integral $\int_{\Sigma}F$ via Stokes' theorem \cite{nakahara}.

The Wilson holonomy may be visualised as the rotation of a ``test arrow" transported around the beam's polarisation texture. For topologically trivial beams, the arrow always returns to its initial orientation, no matter what loop is chosem. Geometrically, this corresponds to a flat connection (i.e., $F=0$), so that $W(\gamma)=\mathbb{I}$. Uniformly polarised, non-vortex beams are the simplest representative of this class. 

When $\mathcal{S}_V[M]=0$, the beam can still have nontrivial 2D or 1D topologies. These can be classified using the reduced Wislon holonomy, obtained by restricting the connection $M$ onto a suitable subspace $F\subset SU(2)$ and calculating $W(\gamma)$ restricted to that subspace via the relation $W_P(\gamma)=P\,W(\gamma)P=\mathcal{P}\,\exp\left(\oint_{\gamma}\,P\,M\,P\right)$, where $P$ is the projector operator that constraints $M$ onto the subspace $F$, in such a way that $M_{2D}$ is a diagonal $SU(2)$ connection for VBs with 2D nontrivial topologies, and $M_{1D}$ is a $U(1)$ (Berry) connection for VBs carrying 1D nontrivial topologies. Skyrmionic beams are an example of the former category. Here,  the arrow twists as it passes around regions of nonuniform polarisation texture, so that particular loops return with nontrivial rotation. In this case, $F\neq 0$ in the textured regions, and $W_P(\gamma)=\exp\left(i\,Q_{2D}\,\uvett{n}\cdot\boldsymbol\sigma\right)$ carries information about the 2D topological charge $4\pi\, Q_{2D}=\int\,d^2x\,\uvett{n}\cdot(\partial_x\uvett{n}\times\partial_y\uvett{n})$, i.e., the skyrmion number. When a beam carries a 1D topological charge $Q_{1D}$, on the ther hand, the arrow acquires a twist as it encircles the defect line generated by a phase or a polarisation singularity. In this case, $F=0$ everywhere except along the singular line (extending in the $z$ direction), and $W(\gamma)=\exp\left(i\,Q_{1D}\varphi\right)$. Vortex beams, beams with isolated polarisation singularities, and VBs that can be mapped on HOPS all fall into this class. For the latter, for example, $2\pi\,Q_{1D}=(\ell-m)$, and $W(\gamma)$ is equivalent to the high-order Pancharatnam-Berry phase introduced in Refs. \cite{milione2,hybridPoincare}. The details of calculation of $Q_{1D}$ and $Q_{2D}$ are given in the supplementary material \cite{supplementary}. More generally, when $F\neq 0$ in a certain volume, the holonomy is genuinely $SU(2)$, i.e., the arrow's orientation depends not only on the path but on its order of traversal. For two distinct loops $\gamma_1$ and $\gamma_2$, for example, the resulting holonomies need not commute in general, i.e., $[W(\gamma_1),W(\gamma_2)]\neq 0$. This is then a direct manifestation of the non-Abelian character of the VB. A pictorial representation of this classification is shown in Fig. \ref{figure3}. This provides a complete classification of the topological properties of VBs in terms of their topological charges and unifies their description in terms of the paraxial bundle. 

\emph{Conclusions -} In this work we have introduced a new theoretical framework for VBs, interpreting them in the language of non-Abelian fibre bundles. By associating a $SU(2)$ connection $M$ to a local polarisation-OAM frame built from orthogonal VBs on distinct HOPS, we uncovered a richer topological structure than the one accessible via their usual Berry phase-HOPS description, in which the Chern-Simons charge and Wilson holonomy emerge as natural invariants to classify VB topologies, generalising and unifying the concepts of winding number, Berry phase, and Skyrmion number for VBs.

This framework establishes an unexplored correspondence between SL and TQFTs, opening a channel for cross-fertilisation of ideas between the two disciplines, that might result in novel ways to control and manipulate the flow of light in (integrated) photonic devices, leading to novel forms of all-optical topological communication and information processing protocols, where the information is encoded and processed in the topological properties of VBs through the use of well-establishes technologies such as $q-$plates and metasurfaces. Extending this framework to include VBs constructed using high-order radial modes, understand how to create a path-integral description of VBs through the connection $M$ and extend these results to the quantum case, thus connecting VBs to the quantum group $SU(2)_k$ \cite{MTCref}, will equip this new framework describing VBs with the necessary tools to actively explpoit the SL-TQFT connection created by this formalism as foundation for topological information and communication protocols based on VBs.

\emph{Acknowledgements -} The author acknowledges the financial support from the Research Council of Finland Flagship Programme (PREIN - decision Grant No. 320165). The author would also like to thank M. R. Dennis for fruitful and insightful discussions.

\end{document}